\documentclass[showpacs,amsmath,amssymb,twocolumn]{revtex4}
\usepackage{graphicx}
\begin{document}

\title{\bf\large{Limit Temperatures for Meson and Diquark Resonances in a Strongly Interacting Quark Matter}}
\author{\normalsize{ Lianyi He, Meng Jin, and Pengfei Zhuang }}
\affiliation{Physics Department, Tsinghua University, Beijing
             100084, China}

\begin{abstract}
We investigate mesons and diquarks as resonant states above chiral
critical temperature $T_c$ in flavor SU(2) Nambu--Jona-Lasinio
model. For each kind of resonance, we solve the pole equation for
the resonant mass in the complex energy plane, and find an
ultimate temperature where the pole starts to disappear. The phase
diagram including these limit temperatures in $T-\mu$ plane is
obtained. The maximum limit temperature at $\mu=0$ is
approximately two times $T_c$.
\end{abstract}


\maketitle

From the collective phenomena observed in high energy heavy ion
collisions at RHIC, the formed new matter is not a weakly coupled
quark-gluon plasma (QGP), but in a strongly coupled
region\cite{shuryak}. Theoretically, the lattice
theory\cite{karsch} and the perturbative resummation
techniques\cite{blaizot, andersen} have told us that QCD matter at
temperatures not far above $T_c$, where $T_c$ is the critical
temperature for deconfinement and chiral restoration phase
transitions, is not a perturbative QGP. Since the interaction is
strong enough, there might be glueballs, mesons, diquarks and
baryons in the system at temperatures $T>T_c$. Recently, the meson
mass above $T_c$ is studied through lattice calculation in chiral
limit\cite{brown}, the multibody bound states are discussed via
variational approach\cite{liao1}, and the lattice baryonic
susceptibility is explained with a picture including baryons at
$T>T_c$\cite{liao2}. If these composite states of quarks and
gluons do exist above $T_c$, they should be melted away when the
temperature of the system is high enough. Therefore, a natural
question one asks is what the ultimate temperature for the
composite states is.

In the idealized case at asymptotically high baryon density, the
phase transitions and the related new matters have been widely
discussed from first principle QCD calculations\cite{highu}. For
the region above but close to the phase transition line, the study
depends on effective models. It is well-known that the
Nambu--Jona-Lasinio (NJL) model\cite{njl} applied to quarks offers
a simple but effective scheme to study chiral symmetry restoration
\cite{chiral}, color
superconductivity\cite{schwarz,huang,shovkovy,ebert,kunihiro,buballa},
and pion superfluidity\cite{barducci,he} at finite temperature and
moderate number densities. In the chiral restoration phase, the
quarks keep massless, and the massive mesons and diquarks can
decay into quark-antiquark and quark-quark pairs. Therefore, the
mesons and diquarks above $T_c$ are not stable bound states, but
rather resonant states, the pole equations for meson and diquark
masses should be regarded in their complex forms in order to
determine the resonant masses and widths self-consistently. This
was done for mesons in the case without considering color
superconductivity\cite{hufner,zhuang}. In this paper, we discuss
the effect of chiral restoration and color superconductivity on
the meson and diquark resonances above $T_c$, and try to extract
the ultimate temperatures of these resonances from the
thermodynamics of the NJL model.

The flavor $SU(2)$ NJL model is defined through the
Lagrangian density,
\begin{eqnarray}
\label{njl}
{\cal L} &=&
\bar{\psi}\left(i\gamma^{\mu}\partial_{\mu}+\mu\gamma_0\right)\psi
+G_s\left(\left(\bar{\psi}\psi\right)^2+\left(\bar{\psi}i\gamma_5\vec{
\tau}\psi\right)^2
\right)\nonumber\\
&+&G_d\left(\bar\psi^c_{i\alpha}
i\gamma^5\epsilon^{ij}\epsilon^{\alpha\beta
\gamma}\psi_{j\beta}\right)\left(\bar\psi_{i\alpha}
i\gamma^5\epsilon^{ij}\epsilon^{\alpha\beta\gamma}\psi^c_{j\beta}\right),
\end{eqnarray}
where $\mu$ is the quark chemical potential, $G_s$ and $G_d$ are
coupling constants in color singlet channel and anti-triplet
channel, $\tau=(\tau_1,\tau_2,\tau_3)$ are Pauli matrices in the
flavor space, and $\epsilon_{ij}$ and
$\epsilon_{\alpha\beta\gamma}$ are totally antisymmetric tensors
in the flavor and color spaces.

\begin{figure}
\begin{center}
\includegraphics[width=6.5cm]{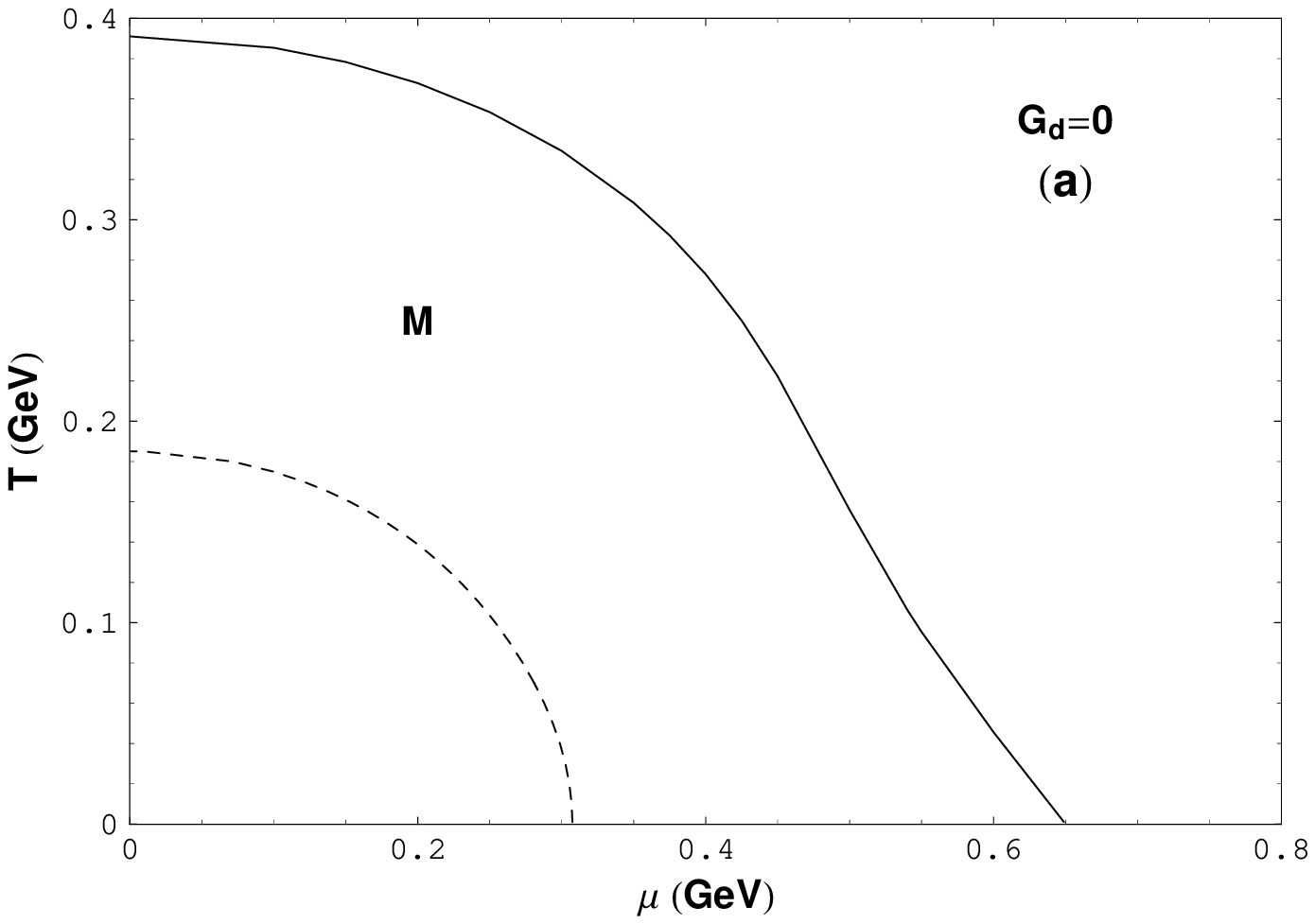}%
\hspace{0.1in}%
\includegraphics[width=6.5cm]{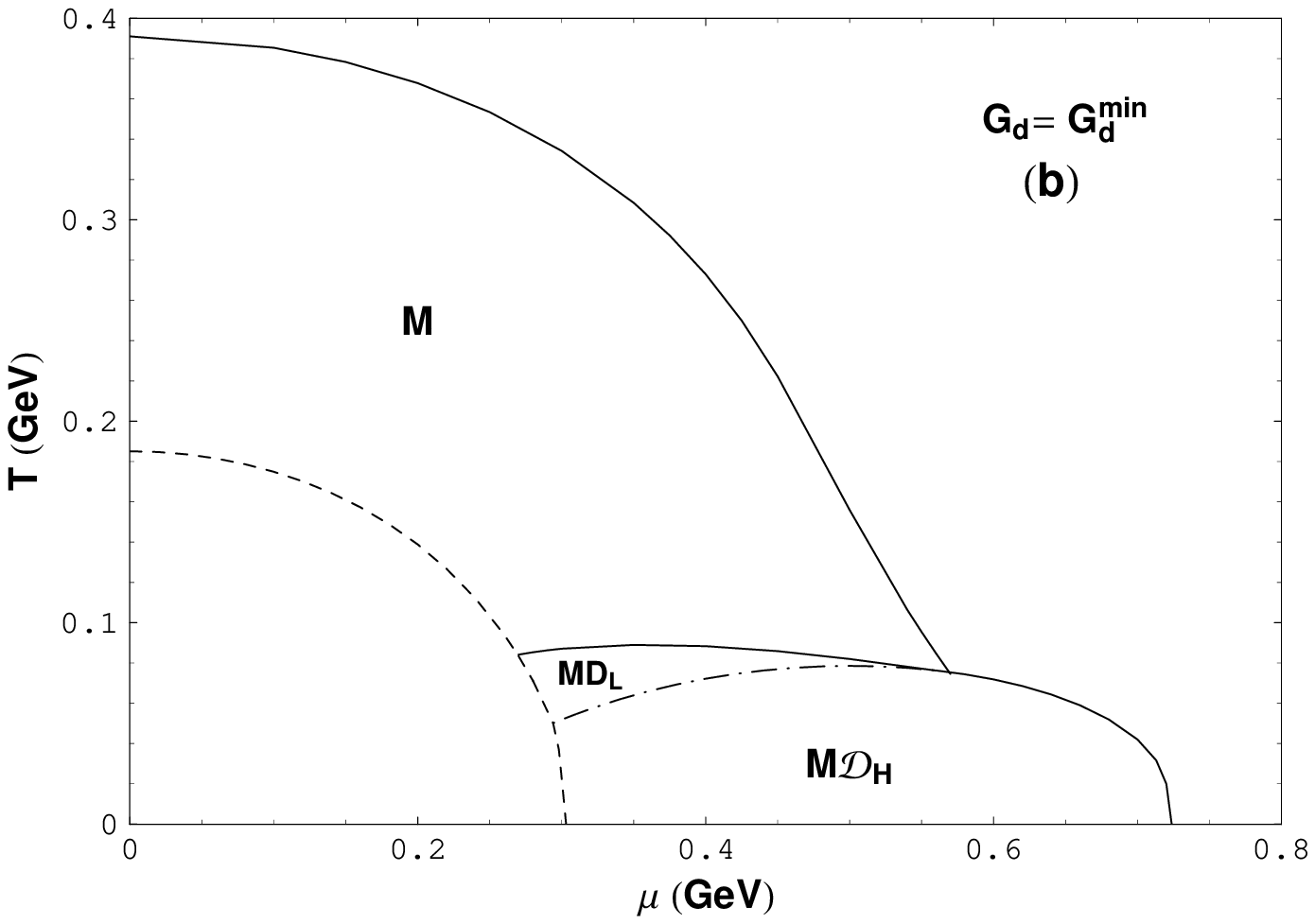}%
\hspace{0.1in}%
\includegraphics[width=6.5cm]{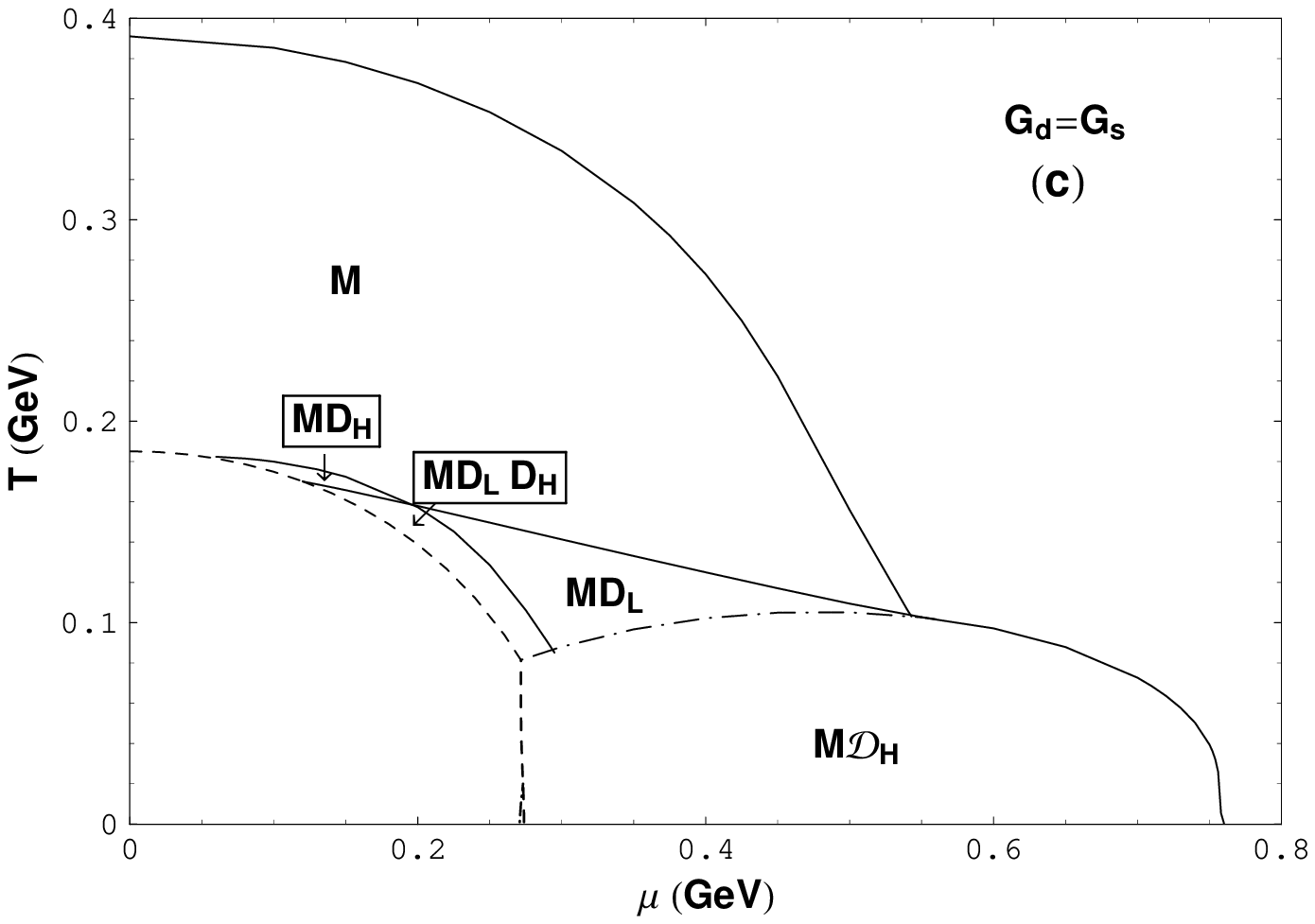}%
\vskip -0.1cm \caption{The critical temperatures for chiral phase
transition (dashed lines) and color superconductivity (dot-dashed
lines) and the limit temperatures (solid lines) for meson and
diquark resonances as functions of chemical potential at coupling
constant $G_d=0$ (a), $G_d^{min}$ (b), and $G_s$ (c). The symbols
$M, MD_L, MD_H$ and $MD_LD_H$ in normal phase indicate the
resonance regions with only mesons, mesons and light diquarks,
mesons and heavy diquarks, and mesons and light and heavy
diquarks, respectively, and $M{\cal D}_H$ means the region of
meson and heavy diquark resonances in color superconductivity
phase. \label{fig1}}
\end{center}
\end{figure}

The quark-antiquark and diquark condensates which are order
parameters of chiral and color superconductivity phase
transitions, respectively, are defined as
\begin{equation}
\label{order}
\sigma = -2G_s\langle\bar\psi\psi\rangle,\ \ \
\Delta = -2G_d\langle
\bar\psi^c_{i\alpha}i\gamma^5\epsilon^{ij}\epsilon^{\alpha\beta
3}\psi_{j\beta}\rangle,
\end{equation}
where it has been regarded that only the first two colors
participate in the diquark condensate, while the third one does
not. In mean field approximation, the two condensates as functions
of $T$ and $\mu$ are determined by the gap equations\cite{huang},
\begin{equation}
\label{gap1}
\sigma\left(1-2G_sI_1\right) = 0,\ \ \ \Delta\left(1-2G_dI_2\right) = 0,
\end{equation}
with the functions $I_1$ and $I_2$ defined as
\begin{eqnarray}
\label{isid}
I_1(\sigma,\Delta) &=& 4\sum_{p,\alpha}{1\over
E_p}\left[{E_p^\alpha\over
E_\Delta^\alpha}\tanh{E_\Delta^\alpha\over 2T}+{1\over
2}\tanh{E_p^\alpha\over 2T}\right],\nonumber\\
I_2(\sigma,\Delta) &=& 4\sum_{p,\alpha}{1\over
E_\Delta^\alpha}\tanh{E_\Delta^\alpha\over 2T},
\end{eqnarray}
where $E^\pm_\Delta = \sqrt{\left(E_p^\pm\right)^2+\Delta^2}$ are
the energies of the effective quarks which participate in the
diquark condensate, $E_p^\pm =E_p\pm\mu$ with $E_p=\sqrt{{\bf
p}^2+M_q^2}$ and quark mass $M_q=\sigma$ are the energies of the
other quarks which are not involved in the diquark condensate, and
$\sum_{p,\alpha}=\int{d^3{\bf p}\over (2\pi)^3}\sum_{\alpha=\pm}$
includes momentum integration and energy summation. The two gap
equations (\ref{gap1}) determine self-consistently the chiral and
color superconductivity phase transition lines in the temperature
and chemical potential plane, which are, respectively, shown as
dashed and dot-dashed lines in Fig.\ref{fig1}. In the following we
focus on the meson and diquark properties above the chiral
critical temperature $T_c$ where $\sigma$ keeps zero, while
$\Delta$ is zero in normal phase but finite in color
superconductivity phase.

In the NJL model, the meson and diquark modes are regarded as the
poles of their effective propagators constructed by the mean field
quarks\cite{chiral,hufner,zhuang}. When the quark propagator in
Nambu-Gorkov space is diagonal in normal phase, the summation of
bubbles in RPA selects its specific channel by choosing at each
stage the same proper polarization function, any of the four
mesons and six diquarks is related to its own polarization
function only\cite{chiral,he,zhuang}. However, when the quark
propagator is with off-diagonal elements in the phase of color
superconductivity, we must consider carefully all possible
channels in the bubble summation in RPA\cite{chiral,he}.

The mesons $\sigma,\pi_+,\pi_-,\pi_0$ are the eigen models of the
system in any phase, the degenerated mass $M_m$ is determined
through the pole equation
\begin{equation}
\label{mass1}
1-2G_s\Pi_{mm}(k_0) = 0,
\end{equation}
with the meson polarization function
\begin{equation}
\label{pim}
\Pi_{mm}(k_0)= I_1+4k_0^2J(k_0),
\end{equation}
where $I_1$ is related to the gap equations (\ref{gap1}) and the
function $J(k_0)$ is given by
\begin{widetext}
\begin{equation}
\label{j}
J(k_0) = -\sum_{p,\alpha}\left[{1\over
2E_p}{\tanh{E_p^\alpha\over 2T}\over k_0^2-4E_p^2}+{E_\Delta^+
E_\Delta^-+E_p^+ E_p^-+\Delta^2\over E_\Delta^+
E_\Delta^-\left(E_\Delta^- + E_\Delta^+\right)}
{\tanh{E_\Delta^\alpha\over 2T}\over k_0^2-\left(E_\Delta^- +
E_\Delta^+\right)^2}+\alpha{E_\Delta^+ E_\Delta^--E_p^+
E_p^--\Delta^2\over E_\Delta^+ E_\Delta^-\left(E_\Delta^- -
E_\Delta^+\right)}{\tanh{E_\Delta^\alpha\over 2T}\over
k_0^2-\left(E_\Delta^- - E_\Delta^+\right)^2}\right].
\end{equation}
\end{widetext}

Since the diquarks condense only in the color $3$ direction, see
the definition of the diquark condensate (\ref{order}), the color
symmetry is spontaneously broken from SU(3) to SU(2) in color
superconductivity phase. While the mass $M_d$ of the four
diquarks $d_1$, $d_2$, $\overline d_1$ and $\overline d_2$,
constructed by the quarks with colors $2$ and $3$, $1$ and $3$,
$\overline 1$ and $\overline 3$, and $\overline 2$ and $\overline
3$, is characterized only by the diquark polarizations
$\Pi_{dd}(k_0)$ and $\Pi_{\overline d\overline d}(k_0)$ in any
phase,
\begin{equation}
\label{mass2}
\left(1-2G_d\Pi_{dd}(k_0)\right)\left(1-2G_d\Pi_{\overline
d\overline d}(k_0)\right) = 0,
\end{equation}
the diquarks $d_3$ and $\overline d_3$, constructed by the quarks
with colors $1$ and $2$, and $\overline 1$ and $\overline 2$, are
no longer the eigen modes of the system in color superconductivity
phase, the new eigen modes are the liner combinations of $d_3$ and
$\overline d_3$, and their mass is controlled by the diagonal and
off-diagonal elements,
\begin{equation}
\label{mass3}
\det\left(\begin{array}{ccc}
1-2G_d\Pi_{d_3d_3}(k_0) &-2G_d\Pi_{d_3\overline d_3}(k_0) \\
-2G_d\Pi_{\overline d_3d_3}(k_0) &1-2G_d\Pi_{\overline d_3\overline d_3}(k_0) \\
\end{array}\right)\ =0,
\end{equation}
with the diquark polarization functions
\begin{eqnarray}
\label{pid}
\Pi_{dd}(k_0) &=&\Pi_{\overline d\overline
d}(-k_0)=I_2-8k_0K_1(k_0)+4k_0^2K_2(k_0),\nonumber\\
\Pi_{d_3d_3}(k_0)&=&\Pi_{\overline d_3\overline
d_3}(-k_0)\nonumber\\
&=&I_2+8k_0K_3(k_0)+4\left(k_0^2-8\Delta^2\right)K_4(k_0),\nonumber\\
\Pi_{d_3\overline d_3}(k_0) &=&\Pi_{\overline
d_3d_3}(k_0)=8\Delta^2K_4(k_0),
\end{eqnarray}
where $I_2$ is related to the gap equations (\ref{gap1}) and the
functions $K_i(k_0)$ are defined as
\begin{widetext}
\begin{eqnarray}
\label{k}
&& K_1(k_0) = -{1\over
4}\sum_{p,\alpha}\alpha\left[\left({E_p^\alpha\over
E_\Delta^\alpha}+1\right){\tanh{E_\Delta^\alpha\over
2T}+\tanh{E_p^\alpha\over 2T}\over
k_0^2-\left(E_p^\alpha+E_\Delta^\alpha\right)^2}
+\left({E_p^\alpha\over
E_\Delta^\alpha}-1\right){\tanh{E_\Delta^\alpha\over
2T}-\tanh{E_p^\alpha\over 2T}\over
k_0^2-\left(E_p^\alpha-E_\Delta^\alpha\right)^2}\right],\nonumber\\
&& K_2(k_0) = -{1\over 2}\sum_{p,\alpha}{1\over
E_\Delta^\alpha}\left[{\tanh{E_\Delta^\alpha\over
2T}+\tanh{E_p^\alpha\over 2T}\over
k_0^2-\left(E_p^\alpha+E_\Delta^\alpha\right)^2}+{\tanh{E_\Delta^\alpha\over
2T}-\tanh{E_p^\alpha\over 2T}\over
k_0^2-\left(E_p^\alpha-E_\Delta^\alpha\right)^2}\right],\nonumber\\
&& K_3(k_0) = -\sum_{p,\alpha}\alpha{E_p^\alpha\over
E_\Delta^\alpha}{\tanh{E_\Delta^\alpha\over 2T}\over
k_0^2-4\left(E_\Delta^\alpha\right)^2},\ \ \ \ \ \ \ \ \ K_4(k_0)
= -\sum_{p,\alpha}{1\over
E_\Delta^\alpha}{\tanh{E_\Delta^\alpha\over 2T}\over
k_0^2-4\left(E_\Delta^\alpha\right)^2}.
\end{eqnarray}
\end{widetext}
It is easy to check that, the pole equation (\ref{mass3}) is
reduced to (\ref{mass2}) for $\Delta=0$ in normal phase.

With the help of the gap equation in color superconductivity
phase, $1-2G_dI_2(0,\Delta)=0$, the mass equation (\ref{mass2})
for the diquarks $d_1, d_2, \overline d_1$ and $\overline d_2$ is
simplified to
\begin{equation}
\label{mass4}
k_0^2\left(k_0^2K_2^2(k_0)-4K_1^2(k_0)\right)=0.
\end{equation}
Obviously, one solution is $k_0^2=0$, and the nonzero root is
determined by
\begin{eqnarray}
\label{mass5}
k_0^2K_2^2(k_0)-4K_1^2(k_0)=0.
\end{eqnarray}

Similarly, the mass equation (\ref{mass3}) for the two new eigen
modes in color superconductivity phase is reduced to
\begin{eqnarray}
\label{mass6}
k_0^2\left((k_0^2-4\Delta^2)K_4^2(k_0)-4K_3^2(k_0)\right)=0.
\end{eqnarray}
Again, one of its solution is $k_0^2=0$, and the nonzero root is
calculated through
\begin{eqnarray}
\label{mass7}
(k_0^2-4\Delta^2)K_4^2(k_0)-4K_3^2(k_0)=0.
\end{eqnarray}
Therefore, there are three massless diquarks and three massive
diquarks in color superconductivity phase. Two of the massive
diquarks determined by (\ref{mass5}) are degenerated and very
light, and the other massive diquark controlled by (\ref{mass7})
is heavy. Taking into account the Higgs mechanism, the three
massless and the double degenerated light diquarks will be eaten
up by gauge fields, only the heavy diquark is left in color
superconductivity phase. In the following, we consider the four
mesons in normal and color superconductivity phases, the six
diquarks in normal phase, and the heavy diquark in color
superconductivity phase.

Above $T_c$ any of the mass equations should be regarded in its
complex form at
\begin{equation}
\label{width}
k_0 = M-i\Gamma/2
\end{equation}
in order to determine the resonant mass $M$ and width $\Gamma$
self-consistently. From the above pole equations, one sees that
this corresponds to solving the conditions
\begin{eqnarray}
\label{widthm}
&& M_m^2-\left(Im{\cal
M}_1(M_m)/2M_m\right)^2-Re{\cal M}_1(M_m)=0,\nonumber\\
&& \Gamma_m=-Im{\cal M}_1(M_m)/M_m
\end{eqnarray}
for the meson mass $M_m$ and width $\Gamma_m$,
\begin{eqnarray}
\label{widthd1}
&& M_d^2-\left(Im\left({\cal M}_2(M_d)\mp
\sqrt{{\cal
M}_3(M_d)}\right)/2M_d\right)^2\nonumber\\
&&\ \ \ \ \ \ \ \ -Re\left({\cal M}_2(M_d)\pm \sqrt{{\cal
M}_3(M_d)}\right)=0,\nonumber\\
&& \Gamma_d=-Im\left({\cal M}_2(M_d)\mp \sqrt{{\cal
M}_3(M_d)}\right)/M_d
\end{eqnarray}
for the masses $M_d$ and widths $\Gamma_d$ of the two triple
degenerated diquarks in normal phase, and
\begin{eqnarray}
\label{widthd2} && M_d^2-\left(Im{\cal M}_4(M_d)/
2M_d\right)^2-Re{\cal
M}_4(M_d)=0,\nonumber\\
&& \Gamma_d=-Im{\cal M}_4(M_d)/M_d
\end{eqnarray}
for the mass $M_d$ and width $\Gamma_d$ of the heavy diquark in
color superconductivity phase, where the complex functions ${\cal
M}_i$ are defined as
\begin{eqnarray}
\label{m}
{\cal M}_1(k_0) &=&
{1-2G_sI_1(0,0)\over 8G_sJ(k_0-i\epsilon)},\nonumber\\
{\cal M}_2(k_0)&=&{1-2G_dI_2(0,0)\over
8G_dK_2(k_0-i\epsilon)}+2{K_1^2(k_0-i\epsilon)\over K_2^2(k_0-i\epsilon)},\nonumber\\
{\cal M}_3(k_0)&=&4{K_1^2(k_0-i\epsilon)\over
K_2^2(k_0-i\epsilon)}\left({K_1^2(k_0-i\epsilon)\over
K_2^2(k_0-i\epsilon)}+{1-2G_dI_2(0,0)\over
8G_dK_2(k_0-i\epsilon)}\right),\nonumber\\
{\cal M}_4(k_0) &=& 4\left({K_3^2(k_0-i\epsilon)\over
K_4^2(k_0-i\epsilon)}+\Delta^2\right).
\end{eqnarray}
The real parts $Re J$ and $Re K_i$ are just the functions $J$ and
$K_i$ themselves, and the imaginary parts $Im J$ and $Im K_i$ can
be written as
\begin{widetext}
\begin{eqnarray}
\label{imagine}
&& Im J(k_0-i\epsilon) =
\pi\sum_{p,\alpha}\left[{\tanh{E_p^\alpha\over 2T}\over
8E_p^2}\delta_1 +{E_\Delta^+ E_\Delta^-+\alpha E_p^+
E_p^-+\alpha\Delta^2\over 2E_\Delta^+ E_\Delta^-\left(E_\Delta^-
+\alpha E_\Delta^+\right)^2}\left(\tanh{E_\Delta^+\over
2T}+\alpha\tanh{E_\Delta^-\over 2T}\right)\left(\delta_\alpha^--\delta_\alpha^+\right)\right],\nonumber\\
&& Im K_1(k_0-i\epsilon) = {\pi\over 8}\sum_{p,\alpha}{1\over
E_\Delta^\alpha}\left[\alpha\left(\tanh{E_\Delta^\alpha\over 2T}
+\tanh{E_p^\alpha\over
2T}\right)\delta_3^+-\left(\tanh{E_\Delta^\alpha\over
2T}-\tanh{E_p^\alpha\over 2T}\right)\delta_3^-\right],\nonumber\\
&& Im K_2(k_0-i\epsilon) = {\pi\over 4}\sum_{p,\alpha}{1\over
E_\Delta^\alpha}\left[{\tanh{E_\Delta^\alpha\over
2T}+\tanh{E_p^\alpha\over 2T}\over E_p^\alpha+E_\Delta^\alpha}
\delta_3^+-{\tanh{E_\Delta^\alpha\over
2T}-\tanh{E_p^\alpha\over 2T}\over E_p^\alpha-E_\Delta^\alpha}\delta_3^-\right],\nonumber\\
&& Im K_3(k_0-i\epsilon) = {\pi\over
4}\sum_{p,\alpha}\alpha{E_p^\alpha\over
\left(E_\Delta^\alpha\right)^2}\tanh{E_\Delta^\alpha\over
2T}\delta_2,\ \ \ \ \ \ \ \ \ \ \ Im K_4(k_0-i\epsilon) =
{\pi\over 4}\sum_{p,\alpha}{\tanh{E_\Delta^\alpha\over 2T}\over
\left(E_\Delta^\alpha\right)^2}\delta_2,
\end{eqnarray}
\end{widetext}
with the $\delta$ functions $\delta_1=\delta(k_0-2E_p),
\delta_2=\delta(k_0-2E_\Delta^\alpha),
\delta_3^\pm=\delta(k_0-(E_p^\alpha\pm E_\Delta^\alpha))$, and $
\delta_\alpha^\pm =\delta(k_0\pm(E_\Delta^++\alpha E_\Delta^-))$
corresponding to energy conservations for different decay
channels. In deriving (\ref{widthm}), (\ref{widthd1}) and
(\ref{widthd2}), we have assumed that the imaginary part $\Gamma$
in ${\cal M}_i$ may be neglected, so that the equations for $M$
and $\Gamma$ are decoupled.

In chiral limit there are three parameters in the NJL model, the
momentum cutoff $\Lambda$ and the two coupling constants $G_s$ and
$G_d$. $\Lambda$ and $G_s$ can be fixed through fitting the
constituent quark mass and pion decay constant in the vacuum,
which leads to $\Lambda=0.65$ GeV and $G_s=5.01$
(GeV)$^{-2}$\cite{zhuang}. While one can not fix $G_d$, we can
determine its low and high limits by taking into account the
physical constraints on the diquark mass in the
vacuum\cite{zhuang1}, $G_d^{min} < G_D < G_d^{max}$ with
$G_d^{min}\sim 0.8G_s$ and $G_d^{max} = 1.5G_s$.

It is now ready to calculate the resonant masses and widths above
the chiral phase transition line. At the critical temperature
$T_c$ where one has $1-2G_sI_1(0,\Delta) = 0$, the pole equation
for the meson mass (\ref{mass1}) is reduced to $ k_0^2J(k_0)=0$,
and the solution is $k_0^2 =0$. Therefore, the degenerated meson
mass evolves with starting value $M_m(T_c)=0$, reflecting
correctly the chiral property at the critical point. From our
numerical calculation, the meson mass goes up monotonously with
increasing temperature, and finally ends at a maximum temperature
$T_r$ where there is no more root for the complex mass equation
and the pole of the meson propagator disappears. This limit
temperature as a function of $\mu$ is plotted as a solid line in
Fig.\ref{fig1}a in the case without considering the diquark
channel. The mesons are in bound states at low temperature
$T<T_c$, but in resonant states in the region $T_c<T<T_r$. The
maximum limit temperature $T_r(0)$ is about two times the chiral
critical value $T_c(0)$. Above the limit temperature $T_r$, the
meson resonances disappear and there are only quarks in the
system.
%%%%%%%%%%%%%%%%%%%%%%%%%%%%%%%%%%%%%%%%%%%%%%%%%%%%%%%%%%%%%%%%%%%%%%%%
%%%%%%%%%%%%%%%%%%%%%%%%%%%%%%%%%%%%%%%%%%%%%%%%%%%%%%%%%%%%%%%%%%%%%%%%

Including the diquark channel, we calculated the meson and diquark
masses and widths as functions of $T$ and $\mu$ for
$G_d=G_d^{min}$ and $G_d=G_s$. For each kind of resonance, there
exists an ultimate temperature where the pole of the corresponding
propagator vanishes, and the resonance disappears from the system.
These limit temperatures as functions of $\mu$ are shown in
Fig.\ref{fig1}b and c as solid lines. The dashed and dot-dashed
lines are the phase transition lines for the chiral restoration
and color superconductivity calculated through the gap equations
(\ref{gap1}). The symbols $M, MD_L, MD_H$ and $MD_LD_H$ in normal
phase indicate the resonant regions with only mesons, mesons and
light diquarks, mesons and heavy diquarks, and mesons and light
and heavy diquarks, respectively, and $M{\cal D}_H$ means the
meson and heavy diquark resonances in color superconductivity
phase.

In normal phase, the limit temperature for mesons is independent
of the coupling $G_d$, while the regions with diquark resonances
depend strongly on the strength of $G_d$. The light triple
degenerated diquarks can be generated at any possible $G_d$, but
the heavy triple degenerated diquarks are created only in the case
with strong coupling. In color superconductivity phase, the heavy
diquark left can be produced everywhere at any coupling $G_d$.
Compared with the case without considering diquark channel shown
in Fig.\ref{fig1}a, the resonant states in Fig.\ref{fig1}b and c
at high baryon density can exist in a wider region, due to the
contribution from the diquark condensate. When the temperature of
the system is higher than the maximum of the limit temperatures,
all the resonances disappear, and there are only quarks in the
system.

In summary, we have investigated the meson and diquark resonances
in a strongly interacting quark matter above but close to the
chiral critical temperature $T_c$ in the NJL model. Above $T_c$,
the massive mesons and diquarks can decay into two massless quarks
and become resonant states.  In mean field approximation to quarks
and random phase approximation to mesons and diquarks, we derived
the pole equations for the resonances in complex energy plane. For
each kind of massive resonances, there exists a limit temperature
$T_r$ where the pole disappears and the resonances are melted in
the hot and dense quark matter. The maximum limit temperature at
$\mu=0$ is approximately two times $T_c$, which agrees well with
the estimated critical temperatures for mesons, diquarks and
baryons in lattice and phenomenological
calculations\cite{shuryak,karsch,brown}. In color
superconductivity phase, the meson and heavy diquark resonances
are survived everywhere.

\vspace{0.3in}

\noindent {\bf \underline{Acknowledgments:}} The work is supported
by the grants NSFC10428510, 10435080 10575058 and
SRFDP20040003103.

\end{document}